\documentclass{article}
\usepackage{graphicx} 
\usepackage{amsmath}
\usepackage{amsfonts}
\usepackage{amssymb}
\usepackage{xcolor}
\usepackage{enumitem}
\providecommand{\keywords}[1]{\textbf{\textit{Keywords:}} #1}

\title{From Mass to Energy-Momentum: \\ The Field-Theoretic Perspective on the Energy-Mass Relation \\ {\small forth. in: \textit{Journal for General Philosophy of Science}}}

\author{Patrick M. Duerr\footnote{Lichtenberg Group for History and Philosophy of Physics, University of Bonn, DE \& History and Philosophy of Science, University of Cambridge, UK; patrick-duerr@gmx.de} \, and J. Brian Pitts\footnote{Mohamed bin Zayed University of Artificial Intelligence, Abu Dhabi, UAE and University of Cambridge, UK; james.pitts@mbzuai.ac.ae, jbp25@cam.ac.uk}}

\date{\today}

\begin{document}

\maketitle
\begin{abstract}
     The paper examines the energy-mass relation of Special Relativity (SR), its status and interpretation, through the lens of classical/non-quantum relativistic field theory. The latter arguably constitutes the fullest embodiment of SR as what Einstein labelled a ``principle theory''. It forms, we propose, the most appropriate perspective also for understanding the energy-mass relation. In a field-theoretic setting, the key notion is a system’s energy-momentum (in local/differential or global/integral form, defined via a suitable energy-stress complex). The energy-mass relation then expresses the role that a system’s \textit{rest}-energy (provided it exists) plays as the functional counterpart of mass: thanks to important theorems, rest-energy qualifies as its field-theoretic generalisation or successor term. That is, while mass is dispensed with as a fundamental notion, rest-energy retains some---but not all---of its salient functional roles; for special cases, it turns out to be directly correlated with mass. In field theory as a general framework for more specific relativistic theories, energy-momentum replaces mass as the essential notion---a profound, yet often overlooked revision in basic physical concepts. Conservation in particular is guaranteed only for energy-momentum, but no longer for mass. In several regards, energy-momentum fuses energy, momentum and mass, with novel connections amongst them and concomitant physical effects. Situating the energy-mass relation within field theory is a surprisingly neglected, but deeply insightful vista for clarifying and philosophically reflecting on SR's foundations. The field-theoretic interpretation achieves a compelling inner coherence and unifying power, and ties it to the rich heuristic resources of field theory, crucial for post-1905 developments in physics.
\end{abstract}

\section{Introduction}

The present paper will revisit what has, quite plausibly, been monikered ``the most famous equation'' of 20\textsuperscript{th} century physics \cite[p. 129]{Lange2001} \cite[p. 62]{Jammer2000}: the energy-mass relation. Its common---albeit historically \textit{in}accurate and systematically misleading \cite{Okun1989,Okun2008,Okun2009,Hecht2009}---form is the iconic $E=mc^2$. Pauli held that ``(w)e may consider [it] (as was done by Einstein) as the most important of the results of the theory of special relativity'' \cite[p. 123]{Pauli}. For Einstein in 1946 the energy-mass relation---both as regards its interpretation and attendant political ramifications---even constitutes ``the most urgent problem of our time'' \cite{EinsteinUrgent}. 

Given this significance of the energy-mass relation, it comes as a surprise that its status and interpretation teem with controversy and confusion (see, \emph{e.g.}, \cite[ch. 3]{Jammer2000} \cite{Fernflores2019}). More worrisomely, extant philosophical reflections on it suffer from three general flaws, all rooted in an unduly narrow theoretical perspective.


First, both derivations and interpretations of the energy-mass relation tend to be confined to particle mechanics (\emph{e.g.}, \cite{Lange2001,Coffey2021ontology,Fletcher2025} for recent examples; cf. also \cite[ch. 3]{Jammer2000} for a more historical perspective). At critical junctures, they all appeal to particle-mechanical notions.\footnote{In fairness, typical physics texts that the philosophical literature draws on are likewise restricted (\emph{e.g.}, \cite[ch 41]{Pauli} \cite[ch. 6]{BergmannBook} \cite[ch. 27]{RIndler1982} \cite[ch. 8]{TaylorWheeler}. The situation is little different for more recent texts (e.g., \cite[Ch.11]{mermin2009} or \cite[Ch.4]{schwarz2010}).} Such a restriction evidently curtails the range of discussions. Moreover, doesn't it verge on the Pickwickian to celebrate the energy-mass relation as the crowning jewel of Special Relativity (SR)---while in the same breath binding it to a theoretical framework that already in 1905 was limping towards obsolescence?\footnote{Einstein \& Infeld depict the ``decline of the mechanical worldview'' as going hand in hand with the rise of classical field theory since Faraday \cite[chs. II, III]{Einstein1938evolution}.}   

Relatedly, this entanglement with particle mechanics stymies the integration of various phenomena for which the validity and meaning of an energy-mass relation appear at first blush opaque: which form does it assume for extended bodies, fluids, or photons (because photons, after all, are generally considered massless but do carry energy)? A systematic approach to an answer seems desirable.

Thirdly and finally, as  \cite[p.124]{janssen2019arches} stresses, SR ``required a `mechanics'---in the sense of a general framework framework for doing physics [...] of \textit{fields} rather than particles''.  The philosophical literature is thus cut off from 
what Einstein, in an unpublished manuscript form 1912, referred to as ``the most important new advance in the theory of relativity'', which ``we owe to the investigations of Minkowski, Abraham, Planck and Laue'' (cited in \cite[p.110]{JanssenElectron}). Physicists usually encounter relativistic (non-quantum) \textit{field} theory as a crucial milestone en route towards its quantum version (\emph{e.g.}, \cite[ch. 1-3]{Maggiore2005qft}). The disconnect belies also the historical relevance of field theory (as well as its enduring significance for neutron star physics, cosmology and heavy-ion collision theory). Laue's (1911) work on continuum mechanics, in particular \cite{LaueContinuum}, paved the way for Einstein's  1913 \textit{Entwurf} theory, Nordstr\"{o}m's scalar theory, and finally General Relativity \cite{NortonNordstrom}. A decisive innovation was the introduction of an energy-stress tensor. In his 1921 Princeton lectures, Einstein  wrote: ``our investigations of the special theory of relativity have shown that in place of the scalar density of matter we have the tensor of energy per unit volume.'' \cite[p. 82]{EinsteinMeaning}.  In Einstein's mature presentation of the field equations a few years earlier, he also underscored the ``energy density tensor'''s relevance for the energy-mass relation: ``the special theory of relativity has led to the conclusion that inert mass is nothing more or less than energy, which finds \textit{its complete mathematical expression} in a symmetrical tensor of second rank, the energy-tensor'' \cite[p. 148, our emphasis]{Einstein1916relativity}. Lamentably, however, philosophical musings on the energy-mass relation, almost without exception, skirt the apparatus of energy-stress tensors.  

The generality and historical fecundity of field theory suggest that we strive to understand the energy-mass relation by the lights of relativistic field theory. As the capstone of ``the most important invention since Newton's time: the field" (as \cite[p. 258]{Einstein1938evolution} put it), we regard such a field-theoretic perspective as an auspicious working hypothesis that this paper will pursue. 

To be sure, the field-theoretic vista prompts deep further questions. In particular: how \textit{exactly} should we construe ``mass'' and ``energy'' for fields?\footnote{Einstein's own ultimately unsatisfactory derivations---according to \cite[p. 591]{Hecht2011}; see also \cite[p. 67 fn. 11]{Jammer2000}), an eyebrow-raising plenitude: ``about 18''!---and interpretation of the energy-mass relation are arguably directly connected to his neglect of the standard (Lagrangian) field-theoretic apparatus (cf. \cite[esp. sect. 3]{EinsteinEnergyStability}). In particular, he seems not to have appreciated the fundamental importance of Noether's work or its antecedents \cite[p. 13]{Okun2008} \cite{BornEnergyMieHerglotz} in grounding conservation laws in rigid symmetries shared by all fields.}

The current paper will attempt to fill these lacunae. We'll unpack and champion an interpretation of the energy-mass relation in terms natural and inherent to classical, special-relativistic field theory. Its main tenets are:

\begin{enumerate}

    \item \textbf{Energy-momentum---not mass---as the \textit{fundamental} notion in relativistic field theory.}
In contrast to Newtonian mechanics (for which mass forms part of the axiomatic foundations), mass no longer figures as a fundamental notion in field theory; its status is more contingent. In field theory, a system's energy-momentum is the more important, natural and fundamental concept. Energy-momentum satisfies conservation laws, but mass does not necessarily.

    \item \textbf{Energy-momentum \textit{unifies} energy and momentum.}
SR entails multiple novel links amongst energy, mass and momentum. Energy and momentum lose their independence; they are amalgamated into energy-momentum---a principal conceptual innovation of relativistic field theory that parallels the unification of electric and magnetic fields in Maxwellian electrodynamics.

    \item \textbf{The energy-mass relation links \textit{rest-energy} and mass.}
    The energy-mass relation, $E=mc^2$, connects energy-momentum and mass in its familiar sense, via rest-energy: 
    \begin{enumerate}[label=(\roman*)]
        \item For time-like energy-momentum (associated with motion slower than light), rest-energy is defined as the energy associated with a system's rest-frame (in which the associated momentum vanishes). It is rest-energy that figures in the energy-mass relation $E_0=mc^2$, not energy \emph{simpliciter}.   
        \item If the system admits of a rest-frame, the Minkowski norm (proper length) of the system's energy-momentum coincides with its rest-energy ($\times c^{-2}$).  
        \item In several theoretical contexts, captured by certain high-level theorems of field theory (to wit: conservation laws in hydrodynamics, Laue's Theorem(s), and the Centre-of-Mass Theorem), rest-energy ($\times c^{-2}$) instantiates several salient functional roles traditionally attributed to (mechanical) mass. Furthermore, for special cases it reduces to mass. It is these functional similarities and special cases that the energy-mass relation encapsulates. ``Mass'' in the energy-mass relation is to be understood primarily \textit{figuratively}, referring to those functional similarities and/or the correlation with mass in special cases.   
        \item In virtue of this functional family resemblance, rest-energy ($\times c^{-2})$ qualifies as the \textit{field-theoretic successor term of mass}---its conceptual generalisation: energy-momentum supersedes the role of mass in the framework of field theory as a fundamental notion, but rest-energy preserves some---but not all---of its previous roles. 
    \end{enumerate}

\end{enumerate}

The \textbf{plan of the paper} is as follows. \textbf{§2} will sketch the basic principles of special-relativistic field theory. Against that backdrop, we'll introduce the notion of energy-momentum and present some key formal results. Equipped with this formal machinery, \textbf{§3} will articulate the field-theoretic interpretation of the energy-mass relation. In \textbf{§4}, we'll examine its chief merits: a cluster of virtues revolving around aspects of coherence and unifying power. \textbf{§5} will recapitulate our arguments and wrap up.

Finally, some limitations of scope are worth stressing at the outset. First, we’ll focus on classical field theory---largely bracketing considerations of quantum fields (except for two remarks \textit{en passant} in order to highlight peculiarities of the classical perspective). Secondly, and more specifically, we’ll focus on field theory on Minkowski spacetime. That is, we’ll be concerned with special-relativistic field theory---bracketing General Relativity (and similar theories of gravity which modify the spacetime background). General-relativistic gravity impinges on the status of energy-momentum, especially energy-momentum ascribable to gravity itself, which raises intricate questions about its definition and interpretation (e.g., \cite{Hoefer,EnergyGravity,DuerrFantasticBeasts,DuerrGravitationalWavesEnergy,DuerrAgainstFunctionalGravitationalEnergy}). Thirdly, we won’t have a lot to say about the boundaries of special-relativistic field theory---what falls within its remit, and what lies outside of it. Interestingly, those boundaries turn out to be vague (see, e.g., \cite{BrownFest,ScalarGravityPhil} for details). Such borderline cases won’t affect our arguments; they hence needn’t concern us here.

\section{The field-theoretical framework}

This section will compile the conceptual tools from special-relativistic field theory that we'll need for shedding light on the energy-mass relation. We'll begin by outlining the general framework of field theory (\textbf{§2.1}), especially but not exclusively relativistic. After briefly clarifying the notion of so-called mass terms in field theory (\textbf{§2.2}), we'll introduce the field-theoretic notion of energy-momentum (\textbf{§2.3}). Together with principal formal results (\textbf{§2.4}), energy-momentum will prove the key concept for understanding the energy-mass relation.


\subsection{Principles of field theory}

Given the iconic status of the energy-mass relation, the present paper aspires to address a philosophically heterogeneous audience. It therefore seems expedient to introduce---at a schematic and non-rigorous level---some elements of field theory not commonly encountered outside a postgraduate physics curriculum (see also \cite{LangePhilPhys} for an accessible introduction pitched to philosophers). Our aim is to isolate those structural features that will prove relevant for interpreting the energy–mass relation.

Field theory describes physical systems in terms of \textit{fields}: quantities defined at each spacetime point (or ``event''), together with their variation across spacetime.\footnote{The ``events” at which fields are defined can also lie in more general---higher-dimensional or more richly structured---spaces (\emph{e.g}., in the $3N$-dimensional configuration space of a system of $N$ particles in quantum mechanics). For simplicity, we confine ourselves to the intuitive case of spacetime.} Mathematically, this involves functions $\phi_K (x^\mu)$, where $x^\mu$ denotes spacetime coordinates and the index $K = 1, \dots, N$ labels the components of the field. A single scalar field (suitable for representing, \emph{e.g}., temperature) has one component, whereas a vector field (suitable for representing, \emph{e.g}., velocities of stellar matter fluid elements) contributes several; more generally, one may consider collections of fields of varying type.

Field theory is motivated by various considerations. Two basic ones stand out:
\begin{enumerate}
\item[(1)] Macroscopically, many phenomena appear as continuous matter distributions: rivers, the atmosphere, steel beams, \textit{etc.}. Hence, it’s imperative to find a mathematical description congenial to modelling bulk matter thus spread-out. Field theory, with fields assigning physical properties to points in spacetime, seems tailor-made for that.
\item[(2)] It’s a plausible and empirically corroborated principle to assume that causal influence doesn’t occur instantaneously, ``at a distance'', but only propagates in a local (usually wave-like) manner. Field theory typically and easily incorporates this requirement, which is otherwise difficult to implement.  
\end{enumerate}

Field theory can be applied at different levels of description. Many physics texts (\emph{e.g.}, \cite{Landau,Maggiore2005qft}) reserve the term for domains where the fields in question are fundamental (\emph{e.g}., as in electromagnetism or the Higgs mechanism), thus excluding continuum mechanics of solids, liquids or gases. But there is also a broader sense of ``field theory’’ that includes continua. It prescinds from the fundamentality of the substratum modelled field-theoretically. Following  \cite[p. 157]{Bunge1967}, \cite{Soper} and \cite[ch. 12]{Goldstein}, we'll adopt this broader perspective. 

Field theory \textit{per se} isn't wedded to specific spacetime symmetries (further attesting to its ``modal cosmopolitanism'' (to recall a phrase \cite{BrownFest})).\footnote{Condensed matter physics provides examples of non-relativistic field theory, \cite{BattermanHydrodynamics}} The theories of interest in this paper, however, namely relativistic ones,  are invariant under the Poincar\'{e} group (comprising spacetime translations, spatial rotations, and Lorentz boosts). These symmetries play a central role: they are closely connected to conservation laws for energy, momentum, and angular momentum (see below, \textbf{§2.3}). By contrast, without necessarily becoming pathological or metaphysically rebarbative (\cite[pp. 548--555]{Goldstein}), field theories lacking such symmetries needn’t exhibit the corresponding conservation laws.\footnote{19\textsuperscript{th} century expectations of a ``uniformity of nature'' may motivate expectations of time- and space-translation invariance (echoed in, \emph{e.g.}, \cite{Landau1976}). But it’s easy to imagine a field theory with explicit dependence on spacetime location: Aristotle's physics, \emph{e.g}., had that feature---hence lacking time- and space-translation symmetries. Anticipating some results from below, in such a field theory conservation of energy and momentum would cease to hold. Another example of such broken time/space invariance that may be of special interest to metaphysicians and philosophers of mind is discussed in (\cite{EnergyMentalCucuLowe} \emph{inter alia}.)} For our purposes, we’ll restrict attention to theories with spacetime translation invariance, and hence to systems in which energy and momentum are conserved (see \textbf{§2.3}).

Relativistic field theories are typically governed by dynamics expressed \emph{via} partial differential equations of a particular technical form, viz. ``hyperbolicity''. It underpins several physically important features. Most notable among them are the finite propagation speed of perturbations and a well-posed initial-value problem (such that ``initial” data across certain surfaces---the equivalents of time-slices---uniquely determines the field values everywhere).\footnote{Hyperbolicity, which comes in various technical flavours, needn’t always be manifest  \textit{directly} in the fundamental equations. In some cases, it emerges only after making conventional choices, such as fixing an electromagnetic gauge or imposing coordinate conditions in General Relativity. The fundamental equations will indicate  the possibility of making such conventional choices.}

The specific dynamics of the fields are most perspicuously formulated via \textit{Hamilton’s principle}: it demands that the so-called action functional be stationary under arbitrary, compactly supported variations of the fields, $\phi \rightarrow \phi +\epsilon \delta \phi $ for some small parameter $\epsilon$. Less formally, one retains only linear terms in $\delta \phi$ or its derivatives. Here, the action $S$ is defined as the spacetime integral of a Lagrangian density $\mathcal{L}$,
\begin{equation}
S[\phi] = \int d^4x \, \mathcal{L}(\phi, \partial_\mu \phi),
\end{equation}
where $\phi$ collectively denotes the fields under consideration and $\partial_{\mu}\phi :=\frac{\partial \phi}{\partial x^{\mu}}$. In a special-relativistic context, it's sometimes convenient to re-write this in terms of the invariant volume element $\sqrt{-\eta}$ induced by the Minkowski metric $\eta_{\mu \nu}$ (which in Cartesian coordinates simplifies to $1$) and a  (scalar) Lagrangian density, though this definition moves a step away from the essence of classical field theory. This move  also tends to obscure instances of the conformal invariance (shape without size) of field theories,  common for ``massless'' fields especially but not only in 4 space-time dimensions \cite{PittsSpinor} (for the pertinent notion of mass for fields, see \textbf{§2.2} below).  


In local relativistic field theories, the Lagrangian $\mathcal{L}$ depends on the fields and a finite number of their spacetime derivatives. In simple cases, $\mathcal{L}$ coincides with the difference between kinetic and potential energy densities. The former typically involves time derivatives of the fields; the latter usually involves spatial derivatives and, in some cases, algebraic terms in the fields themselves (see \textbf{2.2} for these so-called mass terms).

In practice, the construction of Lagrangian densities is guided by a small set of general considerations \cite[ch. 2]{WheelerFieldTheory}. Herein resides the tremendous heuristic power of a Lagrangian approach to field theories. One typically requires that $\mathcal{L}$ be a scalar under the relevant spacetime symmetries (in our case, Lorentz invariance and lack of explicit dependence on spacetime), that it depend locally on the fields and a finite number of their derivatives, and that it respect any additional symmetries taken to characterise the system (such as internal or gauge symmetries). Within these constraints, one usually selects the simplest expressions---only low powers of the fields and their first derivatives---sufficient to capture the phenomena of interest. These principles don’t quite uniquely fix $\mathcal{L}$, but they sharply restrict the space of admissible theories, to be whittled down by further, more context-specific arguments.

Hamilton’s Principle, then, amounts to postulating that the variation of the action vanish:
\begin{equation}
\delta S := \frac{d}{d\epsilon} \Bigl (S[\phi + \epsilon  \delta \phi] \Bigr)|_{\epsilon = 0} = 0.
\end{equation}
This condition yields the \textit{Euler--Lagrange equations}: 
\begin{equation}
\frac{\partial \mathcal{L}}{\partial \phi_K} - \partial_\mu \!\left( \frac{\partial \mathcal{L}}{\partial (\partial_\mu \phi_K)} \right) = 0 \qquad (K = 1, \dots, N).
\end{equation}
They correspond to the equations of motion (i.e., the dynamical equations) for the fields. 

The Euler-Lagrange equations are local conditions: because the variations may be chosen independently within any spacetime region, the resulting field equations must hold pointwise throughout spacetime.\footnote{Appropriate boundary conditions---typically that variations fall off sufficiently rapidly at infinity---ensure that boundary contributions are suppressed in the variational procedure.} 

\subsection{``Mass terms'' in field theory}
For some fields, the Lagrangian contains an algebraic term, quadratic in the field variables, e.g., $-\frac{1}{2}m^2\phi^2$ for a real scalar field (see, e.g., \cite{UnderdeterminationPhoton} and references therein for the analogous case of a mass term for the electromagnetic field). \textit{Conventionally}, it's called a ``mass term''. Yet, this nomenclature isn't arbitrary. 

First, the coefficient $m^2$ has dimensions of inverse length squared (in natural units with $c=\hbar:= h/2\pi \equiv 1$). The parameter $m$ therefore sets an inverse-length scale---or, equivalently, a \textit{mass scale}. 

Secondly, and more importantly, the role of $m$ in the field equations makes clear the connection with the mass in the more familiar sense. For the free scalar, the mass term gives the Klein–Gordon equation $(\Box+m^2)\phi=0$. Its plane-wave solutions obey the dispersion relation $E^2=\mathbf p^2+m^2$, formally analogous to the energy-momentum-mass relation of four-momenta. In cases where the field theory can be quantised, the corresponding excitations are consequently described as particles or quanta of mass $m$. When the quadratic mass term is absent, one obtains the massless relation $E^2=\mathbf p^2$, exactly like for (standard) electromagnetism. 

It's this functional role---rather than any direct identification with a classical mechanical notion of mass---that physicists have in mind when they ascribe a “mass” to a field.\footnote{With an eye to energy-momentum (see \textbf{§2.3}), we underline that although such mass terms \textit{contribute} to a system's energy-momentum, including to its rest-energy (see \textbf{§3.1}), they mustn't be na\"{i}vely equated with the latter.} Foreshadowing a conceptual strategy that we'll deploy later on more systematically (\textbf{§3}), this justification of canonical field-theoretic terminology is similar in spirit to our functionalism about mass.

\subsection{Energy-momentum machinery}

Two results from classical particle mechanics (in its Lagrangian formulation) may serve as a template for treating energy and momentum in field theory: 
\begin{itemize} 
\item Consider the Lagrangian of two masses, $m_1$ and $m_2$, connected by a spring (with spring constant $k$) in free space: $L = \frac{1}{2} m_1 \dot{x}_1^2 + \frac{1}{2} m_2 \dot{x}_2^2 - \frac{k}{2}(x_1 - x_2)^2$. It’s invariant under rigid \textit{spatial} translations $x_i \rightarrow x_i + a$ (for fixed $a$); it depends only on the particles’ relative distance. 
Therefore, the system’s momentum $p_1+p_2$ with $p_i = \frac{\partial L}{\partial \dot{x}_i}$ ($i=1,2$) is conserved, as follows directly from the Euler-Lagrange equations: \begin{equation} \frac{\partial L}{\partial x_i} - \frac{d}{dt} \frac{\partial L}{\partial \dot{x}_i } = 0. \end{equation} 
\item Next, consider the Lagrangian of a particle of mass $m$ in an external, static potential, $L = \frac{m}{2}\dot{x}^2- U(x)$. The Lagrangian doesn’t explicitly depend on \textit{time}. The equations of motion then imply: \begin{equation} \frac{d}{dt} \left(\frac{\partial L}{\partial \dot{x}} \dot{x} - L\right) = - \dot{x} \left(\frac{\partial L}{\partial x} -\frac{d}{dt} \frac{\partial L}{\partial \dot{x}} \right) - \frac{\partial L}{\partial t} = -\frac{\partial L}{\partial t} = 0. \end{equation} 

Consequently, $E:=\frac{\partial L}{\partial \dot{x}} \dot{x} - L $ is conserved. In ordinary cases (as in the example), this corresponds to the system’s energy.\footnote{Energy is frequently introduced as a constant/conserved quantity, suggesting a permanent, putative \textit{substance}. Already at the particle-mechanical level, the Lagrangian perspective hints at a subtle shift away from the primacy, or autonomy, of energy and momentum, and towards a more \textit{structural-relational} emphasis. The latter is then fully enforced, as we’ll see below, in field theory proper: the conserved quantity in question is secondary and derived; instead, what is more primary is a continuity equation, namely a differential conservation law, asserting the absence of energy or momentum \textit{sources/sinks}. The push away from the fundamentality of one unique physical quantity towards the fundamentality of a structural relation, the local conservation law, is further amplified by the occasional ambiguity of a field theory’s Lagrangian, due to the existence of so-called inequivalent Lagrangians (see, \emph{e.g}., \cite{BrownHollandNoether1,SmithConservationExplanation}).} 
\end{itemize} 
These results from particle mechanics generalise. Already in the early 1910s Max Born pointed out how rigid translation symmetries imply conservation laws for fields: \begin{quote} The assumption of [Gustav] Mie just emphasized, that the function [$\mathcal{L}$] is independent of $x,$ $y,$ $z,$ $t,$ is also the real mathematical reason for the validity of the momentum-energy-law. \ldots We assert that for these differential equations, a law, analogous to the energy law (3$^{\prime}$) of Lagrangian mechanics, is always valid as soon as one of the 4 coordinates $x_{\alpha}$ does not appear explicitly in [$\mathcal{L}$]. \cite{BornEnergyMieHerglotz} \end{quote} Both of these links between rigid translation invariance of the Lagrangian and energy/momentum conservation are the simplest instances of Noether’s First Theorem.\footnote{Noether’s contributions from 1918 generalise such results to all rigid symmetries (not necessarily involving spacetime structure), and---via her \textit{Second} Theorem---to local (position-dependent) symmetries (crucial for gauge theories and General Relativity). Finally, often overlooked is her proof of the converses of those results, \emph{viz.} that conservation laws imply symmetries \cite{KosmannSchwarzbachNoether}.} Here, energy and momentum arise as so-called ``Noether charges''.  The conservation of energy due to time translation invariance is closely analogous to the conservation laws for both energy and momentum in field theory.  

In field theory, conservation laws primarily take---in fact, originate in---a local (differential) form. A quantity with density $\rho(t,\mathbf{x})$ and spatial current $J^i(t,\mathbf{x})$ is said to be \emph{locally conserved} iff it satisfies the continuity equation
 \begin{equation} \frac{\partial \rho}{\partial t} + \partial_i J^i = 0, \end{equation} 
where repeated spatial indices $i=1,2,3$ are summed over. This equation expresses the absence of sinks or sources, a balance between the amount of the quantity within a small unit volume and its ambient exterior: a change in the former will be compensated by a flux across the volume’s boundary.
What potential phenomena are excluded?  Nothing can simply disappear \emph{in nihilum}, simply appear \emph{ex nihilo}, or teleport from (say) Abu Dhabi to T\"{u}bingen instantaneously.

Under favourable---but not physically \textit{generic}\footnote{Systems emitting waves, for instance, don’t satisfy the fall-off conditions for global conservation. This is, of course, why such radiative systems aren’t considered isolated or closed: the waves ``escape to infinity''. If we take into consideration present-day cosmology in fact, the transition from local to global conservation laws by integration \textit{fails} in our universe \cite[p. 139]{Peebles}.}---circumstances, a local conservation law allows the definition of an integral quantity which is conserved. We obtain this from integrating over a fixed spatial volume $V$ and using Gauss's theorem: 
\begin{equation} 
\frac{d}{dt} \int_V \rho\, d^3x = - \int_{\partial V} J^i n_i\, dS.
\end{equation} 
If the flux through the boundary vanishes (or decays sufficiently fast \emph{via} suitable fall-off conditions at spatial infinity), one obtains a \emph{global} conservation law for the ``charge'': 
\begin{equation} 
\frac{d}{dt} Q = 0, \qquad Q := \int_{\mathbb{R}^3} \rho\, d^3x . 
\end{equation} 

Global conservation (a Noether charge) is thus the integrated counterpart of a local continuity equation (a Noether current). Still, the local continuity equation is conceptually prior: global conservation is derivative, and conditional on boundary structure. 

Energy and momentum conservation are paradigmatic instances. To see how they emerge as Noether currents, let the dynamics of fields $\phi_K(x)$ ($K=1,\dots,N$) be derived from a Lagrangian density 
\begin{equation} 
\mathcal{L}(\phi_K, \partial_\mu \phi_K, x^\mu), 
\end{equation} 
where $x^\mu=(t,x^i)$ and Greek indices $\mu,\nu=0,1,2,3$. We write $\partial_\mu \phi_K$ also as $\phi_{K,\mu}$. The Euler–Lagrange field equations then read 
\begin{equation} \frac{\partial \mathcal{L}}{\partial \phi_K} - \partial_\mu \frac{\partial \mathcal{L}}{\partial (\partial_\mu \phi_K)} =0 . 
\end{equation}
Suppose now that the Lagrangian density has no explicit dependence on the spacetime coordinates, i.e. $\frac{\partial \mathcal{L}}{\partial x^\nu}=0$ (or, equivalently,  invariance under rigid spacetime translations $x^\mu \rightarrow x^\mu + \varepsilon^\mu$
with constant parameters $\varepsilon^\mu$).

By Noether's First Theorem, each such continuous symmetry gives rise to a conserved current. When subjecting a Lagrangian to Hamilton’s Principle, the variation decomposes into the Euler–Lagrange terms plus a total divergence. Evaluating this for the symmetry-induced variation, one finds---using the field equations (``on-shell”)---that the associated current is locally conserved. For translations, the associated Noether current is the \emph{canonical energy–momentum (or energy-stress) tensor}:\footnote{This expression assumes that $\mathcal{L}$ depends at most on first derivatives of the fields. For most classical field theories of physical interest this is satisfied. If second (or higher) derivatives appear, the Euler–Lagrange equations acquire additional terms.} 
\begin{equation} 
\mathfrak{T}^{\mu}{}_{\nu} = - \mathcal{L}\,\delta^{\mu}{}_{\nu} + \sum_{K=1}^N \frac{\partial \mathcal{L}}{\partial (\partial_\mu \phi_K)} \,\partial_\nu \phi_K . 
\end{equation} 
We use the Gothic letter because the ``canonical tensor'' is actually a weight $1$ density under rigid affine coordinate transformations, assuming that the Lagrangian density $\mathcal{L}$ is taken to be a weight $1$ density to give the invariant action $S = \int d^4x \mathcal{L}$ without introducing potentially surplus structure.  Taking the four-divergence of $\mathfrak{T}^{\mu}{}_{\nu}$ and using the Euler–Lagrange equations yields 
\begin{equation} 
\partial_\mu \mathfrak{T}^{\mu}{}_{\nu} = -\,\frac{\partial \mathcal{L}}{\partial x^\nu}. 
\end{equation} 
Hence, whenever the Lagrangian density is explicitly translation-invariant,
 \begin{equation} 
\partial_\mu \mathfrak{T}^{\mu}{}_{\nu}=0 . 
\end{equation} 
This compact four-dimensional equation comprises four continuity equations: for $\nu=0$ one obtains local conservation of energy, and for $\nu=1,2,3$ local conservation of the three components of momentum. 
Integrating $\mathfrak{T}^{0}{}_{\nu}$ over space yields the conserved four-momentum 
\begin{equation} P_\nu = \int d^3x\, \mathfrak{T}^{0}{}_{\nu}. 
\end{equation} 
Conservation is therefore tied directly to spacetime translation symmetry, just as in the particle-mechanical case, rehearsed above. 
Deferring to convention,\footnote{Note that due to derivation from an expression such as $ \frac{\delta S}{\delta g_{\mu\nu} } \pounds_{\xi}g_{\mu\nu} = 2 \frac{\delta S}{\delta g_{\mu\nu}} \nabla_{\mu} \xi_{\nu},$ even the Hilbert-Rosenfeld stress-energy definition naturally begins as a mixed tensor density of weight $1$ before one de-densitises it and moves an index with the metric! A mixed weight $1$ tensor density is just what one would want to turn a tangent (``contravariant'') vector describing a translation into a tangent vector density of weight $1$ as a current, the kind of entity for which the covariant derivative is the same as the partial derivative, hence optimal for integration.} henceforth we will use a non-weighted energy-momentum tensor $T$ with some indices but no Gothic font; the relation is ${T}^{\mu}{}_{\nu}=_{def} \frac{1}{\sqrt{-\eta} } \mathfrak{T}^{\mu}{}_{\nu}$.  

Let’s cap off our \textit{tour d’horizon} on field-theoretic energy-momentum with some comments on energy-momentum currents. A warning against their na\"{i}ve reification is in order. The notion of energy-momentum \textit{isn’t unique}: one can change a stress-energy tensor by a quantity with automatically vanishing divergence, getting another stress-energy tensor ascribing different amounts of energy per unit volume, \emph{etc.} Less worrisomely, one can also add terms proportional to the field equations or their derivatives, which have a value of $0.$

A common attitude towards the canonical energy-momentum $T^{\mu}{}_{\nu}$ is not to regard it as the last word, but a stepping stone towards a more satisfactory notion. Two key vices are usually adduced (\cite{ForgerRomerStress}).
First, canonical energy-momentum fails to be gauge invariant. For instance, under gauge transformations $A_\mu \rightarrow A_\mu +\partial_\mu \Lambda$, canonical energy-momentum for electromagnetic fields acquires an additional term that depends on the choice of $\Lambda$. A related point is that the  canonical energy-momentum is sensitive to the choice of Lagrangian: while dynamically inconsequential, a term added to the Lagrangian density that doesn't affect the Euler-Lagrange equations does change the canonical energy-momentum by a divergence-free term. 

Secondly, canonical energy-momentum isn’t, in general, symmetric: for a scalar field it is; for instance, for electromagnetism, by contrast, it isn’t.\footnote{With one index up and one down, the question of symmetry isn’t even well-posed unless one has at one’s disposal a spacetime metric $\eta_{\mu\nu}$ with which to raise or lower indices. In relativistic field theory, such a metric is present, of course.} This fact is often taken to signal a defect for two reasons.\footnote{One may impugn the force of these reasons. The first trades on a particular definition of angular momentum density, $M^{\mu\nu\alpha} = x^\mu T^{\nu\alpha} - x^\nu T^{\mu\alpha}$. Its conservation indeed demands $T^{\mu\nu} = T^{\nu\mu}$. However, this definition is incomplete for fields carrying non-zero spin \cite[pp. 111]{Soper}. When the \textit{full} Noether current associated with Lorentz invariance is computed, angular momentum conservation follows \textit{without} requiring the canonical tensor itself to be symmetric. 
The second reason invokes General Relativity's symmetric ``Hilbertian'' energy-momentum tensor, defined \textit{as the variation of the matter action with respect to the metric}---a distinct object  (\cite[pp. 236, 269-271]{WeylSTM} \cite[p. 280]{Landau} \cite[p. 465] {MTW} for salient differences).  In General Relativity, one thus doesn’t so much \textit{repair} canonical energy-momentum as \textit{replace} it.} One is the claim that angular momentum conservation requires a symmetric energy-momentum expression; the other is the claim that coupling matter to gravity (in the context of General Relativity) requires a symmetric expression for energy-momentum. 

Vis-\`{a}-vis these drawbacks one may wish to ``improve'' canonical energy-momentum. Two such routes are the systematic procedure due to Belinfante (\cite[pp. 114-121]{Soper}), or the modification of Hilbert’s variational definition of energy-momentum, due to Rosenfeld \cite{RosenfeldStress,Deser,PonsEnergy}.\footnote{To the best of our knowledge, no systematic comparison of the advantages and disadvantages of the various proposals is forthcoming, nor a sustained discussion and analysis of desiderata for energy-momentum proposals.}

Here, we forgo plumbing the depths of this quandary. What matters for our purposes is the system’s globally defined energy-momentum, which, given suitable boundary conditions\footnote{Boundary conditions at spatial infinity might seem too lax, however.  There are some further, loose constraints:  for our purposes one wouldn't want to permit the ``relocalisation'' of the energy of some object to be quite outside the object's rough-and-ready location.}, 
remains unaffected. It’s this quantity that enters the field-theoretic interpretation of the energy-mass relation.

\subsection{Key theorems}

Central to the field-theoretic underpinning of the energy-mass relation are two theorems: Laue’s Theorem (\textbf{§2.3.1}), and the Centre-of-Mass Theorem (\textbf{§2.3.2}). What makes them especially interesting for our purposes is their abstractness: they don’t rely on specific dynamics or other details from (relativistic) matter theories (see also \textbf{§3.2}).

\subsubsection {Laue’s Theorem} 

It’s apposite to discriminate between two versions of Laue’s Theorem: a passive (``Laue-Klein’’) and an active (``Laue-Giulini’’) formulation, where the former deals with active Lorentz transformations (which, without changing coordinate systems or reference frames, describe otherwise identical systems that move with uniform speed or are spatially rotated), and the latter with passive ones (which describe the same event or object using different coordinates or reference frames).
The objects of both are time-independent/stationary and closed/isolated systems. That is, for these systems, local energy-momentum conservation, $\partial_\nu T^{\mu \nu} = 0$, and stationarity, $\partial_0 T^{\mu \nu} = 0$, hold alongside suitable boundary conditions (forestalling energy-momentum leakage into infinity).

\textit{The Laue-Klein\footnote{Klein generalised Laue's derivation to allow for closed but time-dependent systems \cite{Ohanian2012} (with his proof being recalled in Pauli's and Weyl's books \cite[pp. 60-61, 87, 125]{Pauli} \cite[pp. 271-273, 323, 324]{WeylSTM}).} Theorem}, the historically first version \cite[section 9.3 especially]{NortonNordstrom} (see \cite{Ohanian2009,LaueTheoremWang, Ohanian2012} for technical and historical details), deals with energy-momentum, defined component-wise for a given observer with the simultaneity hypersurface $\Sigma = \mathbf{n}^\perp$ (and the associated unit normal $\mathbf{n} = (n^\mu)$):

\begin{equation}
P^\mu [T, \Sigma] := \int_\Sigma d^3 x \, \sqrt{-\eta} \, T^\mu_{\ \nu} \, n^\nu.
\label{LAUE-KLEIN}
\end{equation}

Here, we used the 3-volume element $\sqrt{-\eta}d^3x$ (induced by the Minkowski metric on $\Sigma$) for greater generality (to allow for non-Cartesian coordinates). 

Note that we integrate some frame-dependent components of a tensor over space, which is in turn a frame-dependent operation (due to the relativity of simultaneity). One may hence wonder whether $P^\mu$ is a well-defined object---whether it represents a covariant quantity (a $4$-vector) or is rather irredeemably tainted with descriptive artifacts.

Laue’s Theorem answers this in the affirmative. More precisely, it shows that if the energy-momentum tensor satisfies suitable fall-off conditions at infinity and if the integrals in (\ref{LAUE-KLEIN}) converge (are finite), then the energy-momentum forms a well-defined geometric object of the right type: 
\begin{equation}
P'^\mu = P^\mu [\Lambda \cdot T, \Lambda \cdot \Sigma] = \Lambda^\mu _{\ \nu} P^\nu [T, \Sigma].
\end{equation}
In other words, under a \textit{change of coordinates}, related via a Lorentz transformation represented by the matrix $\Lambda = (\Lambda^\mu _{\ \nu})$, the $P^\mu$ form a four-vector, the energy momentum $\mathbf{P} = (P^\mu)$, independent of the choice of simultaneity slice: $P^\mu$ and $P'^\mu$ are components of the same (coordinate-independent) object $\mathbf{P}$. The isolated and stationary system’s energy-momentum, Eq. \ref{LAUE-KLEIN}, thus seamlessly fits into SR's four-dimensional Minkowskian ontology (see, \emph{e.g.}, \cite{nerlich2013einsteins,Coffey2021ontology}), and behaves like the energy-momentum in relativistic mechanics and electrodynamics. 

In \textbf{§3.1.1}, we'll see how, thanks to a simple corollary, the Laue-Klein Theorem grounds an even closer structural similarity between the energy-momentum and the energy-momentum of a point-particle.

For the \textit{Laue-Giulini Theorem}---a variant of the Laue Theorem, due to \cite{Giulini2018}---one interprets the Lorentz transformations \textit{actively}. It shows that for stationary and closed systems whose energy-momentum tensor falls off sufficiently at infinity, an active Lorentz boost of a system results in a Lorentz boost of the energy-momentum:

\begin{equation}
P^{\mu}[\Lambda\!\cdot\!T,\Sigma]
 \mathrel{\mathop{=}\limits^{\text{def}}} \int_{\Sigma} d^{3}x\;\sqrt{-\eta}\;
\Lambda^{\mu}{}_{\kappa}\,\Lambda^{\lambda}{}_{\nu}\;
T^{\kappa}{}_{\lambda}\;n^{\nu}
=
\Lambda^{\mu}{}_{\nu}\;P^{\nu}[T,\Sigma]\,.
\end{equation}

Note that in contrast to the Laue-Klein Theorem, the boost effects a \textit{distinct} object: $\mathbf{P} = (P^\mu) \not \equiv \mathbf{P'} = (P'^\mu) = \Lambda \cdot \mathbf{P}$. The two energy-momenta, however, have the same rest-energy.

In \textbf{§3.1.1}, we'll see how, thanks to simple corollaries, the Laue-Giulini Theorem strengthens the interpretative implication that the Laue-Klein version entailed for energy-momentum Eq. (\ref{LAUE-KLEIN}).

\subsubsection{Centre-of-Mass Theorem}
As foreshadowed in the mechanical case (\emph{e.g}., \cite[ch. 8]{Landau1976}), the general idea of a centre-of-mass motion theorem requires little more than the relativity principle (of the Galilean or relativistic variety) to ensure the conservation of the requisite additional quantities besides energy, momentum, and angular momentum. 

Given the (in our experience) unfortunate relative unfamiliarity of this theorem, we sketch its provenance (see \cite[Ch.9]{Soper} and Wheeler \cite[ch. 2]{WheelerFieldTheory} for details). Start from the definition of angular momentum density 
as \begin{equation} J^{\alpha\beta\mu} = 
x^{\alpha} T^{\beta\mu} - x^{\beta} T^{\alpha\mu}
\end{equation}
for a symmetric energy-momentum  tensor (see \textbf{§2.3}). Next, integrate the conservation equation for angular momentum, $\partial_{\mu} J^{\alpha\beta\mu} = 0$, over the simultaneity surface $\Sigma = \{t=const.\}= \mathbf{n}^\perp$ with the associated unit normal $\mathbf{n}$), discarding a spatial boundary term (assuming suitable boundary/fall-off conditions):
\begin{equation}
 \frac{d}{dt} \int_{\Sigma} d^3x \sqrt{-\eta} (x^\alpha T^{\beta \gamma}- x^\beta T^{\alpha \gamma})n_\gamma = 0. 
\end{equation}
The expression being anti-symmetric in ($\alpha, \beta)$, the diagonal components ($\alpha=\beta$) vanish. While the non-diagonal, purely spatial indices (i.e., $0\neq\beta\neq\alpha \neq 0$) yield an angular momentum density, for our purposes the non-diagonal, spatiotemporally mixed ones (i.e., $\alpha= 0$  and $\beta := m =1,2,3$) are of interest: 
\begin{equation} 
0=\frac{d}{dt} \int d^3x \sqrt{-\eta}  (x^{0} T^{m \gamma} - x^{m} T^{0 \gamma})n_\gamma.
\label{CENTRE}
\end{equation}
Noting that energy-momentum $(P^\mu)=(E/c, P^m)$ is conserved, the last term in Eq. (\ref{CENTRE}) can be interpreted as the position $X^m (t) $ of the centre-of-energy times the total conserved energy ($\times c^{-2}$), $E/c^2$. Thus,
\begin{equation} 
0=\frac{d}{dt} (ct P^m - \frac{E}{c} X^m) =cP^m - \frac{E}{c} \frac{dX^m}{dt}.
\label{CENTRE-OF-MASS}
\end{equation}
In conclusion, we find that the centre-of-energy moves rectilinearly, with constant velocity $d X^m (t)/dt$. In \textbf{§3.1.4}, we revert to Eq. (\ref{CENTRE-OF-MASS}), for an even closer link to mass in particle-mechanics \emph{via} the velocity-momentum relation.   

The foregoing expression is somewhat compromised by its plain non-covariance. This can be remedied \cite[sect.3]{Lorce} by introducing a (non-unique) time-like, future-directed covector field. It represents possible observers in a reference frame (to which the coordinate system  $(t,X^m)$ is \textit{not} necessarily adapted). Although different observers locate the centre-of-energy at different locations (as one would expect), in the system’s rest-frame of the energy-momentum, they all coincide (op.cit., p.6). 


\section{The field-theoretic view on the energy-mass relation}

Drawing on the formal results surveyed in \textbf{§2}, we'll now expound the interpretation of the energy-mass relation as it naturally arises within relativistic field theory. \textbf{§3.1} will delineate its core tenets. \textbf{§3.2} will spell out further details to elucidate the significance of those tenets and their ramifications. 

\subsection{Core tenets}

Our proposed field-theoretic interpretation of the special-relativistic energy-mass relation comprises two main tenets: (ELIMINATIVISM) and (MASS FUNCTIONALISM), respectively. 

The first clarifies the \textit{status} that the concept of mass enjoys---and contrariwise \textit{doesn't} enjoy---in field theory. The second clarifies the links---or \textit{functional similarities}---between mass, as traditionally understood, and the concept that supersedes it, energy-momentum.

\textbf{(ELIMINATIVISM)} encompasses three assertions, expressing the conceptual revisions that field theory inaugurates \textit{vis-à-vis} particle and continuum mechanics:

\begin{itemize}
    
  \item In pre-relativistic or relativistic \textit{mechanics}, mass (or mass density) is a conceptual-axiomatic prerequisite (see, e.g., [Ch.3]\cite{Bunge1967}); physically, mass (or mass-density) represents intrinsic, essential properties of matter (see, e.g., [Ch.1--2]\cite{Jammer2000}). In field theory, mass no longer figures as an indispensable, or even central posit in either sense; the foundations of field theory don't presuppose any notion of mass, neither structurally nor ontologically. A physical entity needn't be ascribed mass; whether it does (as in the case of, say, fluids) or doesn't (as in the case of, say, gravitational or electromagnetic fields) possess mass in a sufficiently clear, inherent sense is, from the perspective of field theory, an accidental feature of the system one describes field-theoretically---not a physical or metaphysical necessity.  
 
  \item The central object is instead energy-momentum (\textbf{§2.3}), a \emph{sui generis} quantity.  
  \item Mass conservation is no longer a separate postulate. Its validity isn't presumed any more: whether it holds or not is relegated to matter theories---beyond the ambit of field theory as a \textit{framework} for more specific interaction theories (cf. \cite{Flores1999-FLOETO}). What field theory does supply (subject to physically reasonable conditions) is the conservation of energy-momentum, in both local and global form (\textbf{§2.3}).
\end{itemize}

Having stripped mass of its received status as a foundational notion, how does field theory provide an interpretation of the mass-energy relation? Here, (MASS FUNCTIONALISM) enters the stage. It spells out how mass, as traditionally conceived, and energy-momentum are connected. We'll now expand successively on the general decomposition of energy-momentum into energy and momentum, the link between rest-energy and mass more specifically, and the more general link between energy-momentum and mass. 

In field theory, a time-like energy-momentum $\mathbf{P}$ admits of a unique decomposition into parts, parallel and orthogonal, to a time-like observer $\boldsymbol{\xi} = (\xi^\mu)$ (see, \emph{e.g.}, \cite[ch. 9]{Gourgoulhon2013}): 

\begin{equation}
   \mathbf{P} \equiv \frac{1}{c^2}(\mathbf{P} \cdot \boldsymbol{\xi}) \,\boldsymbol{\xi} 
   + \big(\mathbf{P} - \frac{1}{c^2}(\mathbf{P} \cdot \boldsymbol{\xi}) \,\boldsymbol{\xi}\big) 
   = (E[\xi]/c) \,\boldsymbol{\xi} + \mathbf{p}[\xi], 
   \label{DECOMPOSE}
\end{equation}
with the energy $E[\xi]/c := \frac{1}{c^2}\mathbf{P} \cdot \boldsymbol{\xi} = \frac{1}{c^2} \eta_{\mu\nu} P^{\mu} \xi^{\nu}$ relative to (projected onto) the observer, and the three-momentum $\mathbf{p}[\xi] := \mathbf{P} - E[\xi]\boldsymbol{\xi}$, as the orthogonal complement ($\mathbf{p} \perp \boldsymbol{\xi}$). Here $\eta_{\mu\nu}$ is the spacetime metric tensor; we use signature $+---$.  For many purposes it suffices to use Cartesian coordinates, setting $\eta_{\mu\nu} = diag(1,-1,-1,-1).$ 


Thanks to the orthogonality of $\mathbf{p}$ and $\boldsymbol{\xi}$, the energy-momentum's norm square decomposes into the sum of both:
\begin{equation}
   \mathbf{P}^2 \equiv E^2/c^2 - p^2. 
   \label{MOMENTUM-ENERGY-3momentum}
\end{equation}   

For a time-like energy-momentum there always exists an observer $\boldsymbol{\xi}_0$ such that the associated 3-momentum vanishes: $(\mathbf{P} \cdot \boldsymbol{\xi}_0)\boldsymbol{\xi}_0 - \mathbf{P} = 0 $. Thus $\boldsymbol{\xi}_0$ is parallel to $\mathbf{P}$. The corresponding energy $E_0 := E[\xi_0] := \mathbf{P} \cdot \boldsymbol{\xi}_0$ is called the ``\textit{rest-energy}''.  
Consequently, we have the identity:
\begin{equation}
   c^2\mathcal{M}^2 := \mathbf{P}^2 = E_0^2/c^2. 
   \label{NORM}
\end{equation}   

The evocative\footnote{Note that, \emph{e.g.}, (\cite[p. 272]{Gourgoulhon2013}) \textit{defines} mass as the energy-momentum's norm. We reject this: according to (MASS FUNCTIONALISM), $\mathcal{M}$ displays merely a functional family resemblance with mass.} symbol for this definition calls for a justification: how is $\mathcal{M}$, and in virtue of eq. (\ref{NORM}) also $E_0$, related to mass in the ordinary sense, familiar from mechanics?
Our strategy for an answer is broadly functionalist\footnote{This approach to counterparts of a term, defined in one theoretical context, for a different theoretical context, \emph{via} principal functional roles follows a recent trend in philosophy of physics \cite[p. 58]{WallaceEverett} \cite{ReadEnergy,Lam2020-LAMSIA-3}. Underlying (MASS FUNCTIONALISM) is something close to what Knox \& Wallace \cite{KnoxWallace}  call ``constitutive functionalism'' (as opposed to ``causal-role functionalism''). 

Given the polysemy of ``mass'' \cite{Hecht2006}---arguably characteristic of many concepts in science---a family resemblance approach seems especially natural.}: under certain---but \textit{not} generic---circumstances, it instantiates structural relations distinctive of mass as it figures in (classical or relativistic) mechanics; it approximates the mass role under those conditions. In virtue of this family resemblance as regards the functional profile of (mechanical) mass, rest-energy ($\times c^{-2}$) counts as a generalised notion of mass in field theory. We'll dub this rationale \textbf{(MASS FUNCTIONALISM)}. It adverts to four different contexts in which the mass role of rest-energy is instantiated.  
\subsubsection{Limiting cases}
First, rest-energy is firmly tethered to mass in certain limits.\footnote{Interestingly, according to \cite[p.110]{JanssenElectron}, the reconstruction of particle mechanics from a continuum mechanics of fields ``was first worked out explicitly in the context of general rather than special relativity'' by Einstein and Klein in 1918. This fits the fate of special-relativistic field theory being repeatedly sidelined in both history and philosophy of physics (see also op.cit., pp.67-68).} A \textit{direct computation} (or evaluation) of the Noetherian/canonical energy-momentum tensor, and its associated integral/global quantities, for specific physical systems with (mechanical) mass yields an energy-momentum whose norm square is $m^2 c^2$; here $m$ corresponds to the familiar inertial (or rest\footnote{Instead of ``rest mass'' we'll stick to ``mass'' simpliciter for two reasons. First, mass is traditionally conceived of as an intrinsic property, ``the essential (because unvarying) quality of matter'' \cite{EinsteinUrgent}. Secondly, and relatedly, to speak of ``rest mass'' suggests \textit{non-rest} forms of mass. Historically, these have been the so-called ``longitudinal'' and ``transversal'' masses (see, \emph{e.g.}, \cite[p. 82]{Pauli}). Their legitimacy and usefulness are, however, controversial (\emph{e.g.}, \cite[Ch. 2]{Jammer2000} \cite{BeisbartJung,Hecht2009,fliessbach2018relativistische}.}) mass from classical and relativistic mechanics. For a point-particle of mass $m$---either free, or subject to velocity-independent interactions that only modify the potential energy (and result in external forces)---one straightforwardly verifies this.

A more general case to the same effect concerns the perfect fluid. Expanding the relativistic Euler Equation in an external field (an immediate consequence of $\mathbf{\nabla} \cdot \mathbf{T} = 0$) in orders of relativistic correction factors, one gleans that in leading order (the so-called ``weakly relativistic limit'' of relativistic hydrodynamics, see, \emph{e.g.}, \cite[ch. 3]{rezzolla2013})---when rest mass energy dominates, as in ordinary matter or interstellar gas at low temperatures---rest-energy and mass ($\times c^2$) of the fluid coincide.     
Together with eq. (\ref{NORM}), we thus get $\mathcal{M} = m$. For these special cases, the (rest) energy-mass relation $E_0 = mc^2$ is reproduced exactly. 
In pre-relativistic field theory, no such inherent correlation between (rest) energy and mass holds. In relativistic field theory, the inherent correlation is a structural conceptual novelty, emanating from its basic principles. 
     
The expression $mc^2$ can be understood as an ``internal'' (or, in Einstein's (1919) words: ``latent'' \cite[p. 230]{Einstein1919Times}) contribution to the total energy. In virtue of the principles of field theory, the mass of the system, as conceived of in mechanical terms, is directly correlated with its energy content. While the foregoing limiting cases provide an important conceptual link, the rubber meets the road physically-empirically in the next context.

\subsubsection{Energy and momentum conservation in fluids}

(MASS FUNCTIONALISM) points to the functional role of mass concerns energy and conservation laws for perfect---but \textit{pressureless}\footnote{For our present purposes, it's crucial that the pressure remains negligible. Else, we get an effective inertial mass density to which also pressure contributes (see, \emph{e.g.}, \cite[p. 668]{Gourgoulhon2013}). We'll revert to this later on when making the more general case that energy-momentum is most naturally seen as the successor term of mass, which supersedes the latter, but remains intimately connected via rest-energy.}---fluids. Quotidian fluids and gases (\emph{e.g.} air or water), but also ordinary solids (\emph{e.g.}, crystals) can be modelled approximately in this way, at non-extreme temperatures.

The relativistic Euler Equations (also in the presence of an external energy or momentum sink) then yield statements of energy and momentum conservation where the rest-energy density $\varepsilon_0:= T^{\mu \nu} u_\mu u_\nu$ (up to $c^{-2}$) plays the role of the inertial mass density in the pre-relativistic case. The energy balance takes the form: 
\begin{equation}
(\varepsilon_0/c^2) a^\mu = \text{external force},
\end{equation} where  
$
a^\mu = u^\nu \nabla_\nu u^\mu$ denotes the four-acceleration of a comoving, infinitesimal fluid element (with the four-velocity $u^\mu$). The evolution of the fluid element's momentum obeys the continuity equation
\begin{equation}
\nabla_\mu \left((\varepsilon_0/c^2) u^\mu \right) = \text{energy sinks/sources}.
\end{equation}
The covariant derivative $\nabla_{\mu}$ (defined via the Minkowski metric's Levi-Civita connection) is useful if one doesn't use Cartesian coordinates (as is common for fluids).  

Since the rest-energy density $\varepsilon_0/c^2$ plays this double role of inertial mass density, the associated integral quantity, rest-energy $E_0$, inherits this link: $\mathcal M$  plays the role of inertial mass in the relativistic dynamics of systems that can be modelled as perfect, pressureless fluids. 

Note that this jolts the previously established correlation into predictively novel territory: it implies that increasing or decreasing the rest-energy (\emph{e.g.}, by absorbing or emitting photons) alters the effective inertial mass. Heating up a box filled with a diluted gas (and hence negligible pressure), for instance, will increase the effective inertial mass, as determined by the ratio of bulk acceleration and external force of that box.\footnote{An instructive, similar case is discussed by \cite[p. 320]{lanczos1970}: the consistent treatment of a massive particle in a scalar potential $\Phi(x)$. This leads to an effective inertial mass $m+\Phi/c^2$, precisely the rest energy $\times c^{-2}$, in the equations of motion: ``(i)f we wind up a watch, thus increasing its latent potential energy by a certain amount of elastic energy we change the mass of the watch'' by the amount $\Phi/c^2$. ``If a certain chemical compound changes its configuration [...]---thus giving off the corresponding amount of chemical energy in the form of heat---the mass of the new compound has diminished'' (op.cit., p.321). NB: This result follows from very general considerations of ``(harmonisation) with the principle of relativity'' (op.cit., p. 320). Lanczos also gives an astute geometric explanation for why we \textit{don't} see a similar dependence of effective mass on the electromagnetic four-potential (op.cit., p. 327).} 

\subsubsection{Laue's Theorem(s)}

A third context stems from a philosophically---unfairly---neglected corner: Laue's Theorem in its two formulations from \textbf{§2.3.1}.

According to the \textit{Laue-Klein Theorem}, for $\mathbf{P}^2 >0$\footnote{In the general-relativistic context, this ensured through so-called energy conditions (see, e.g., \cite{CurielEnergyConditions}).}, the energy-momentum \textit{of an arbitrary system} (subject to the conditions of the Theorem's validity) resembles a relativistic point-particle's with the mass $E_0/c^2$; both share the same form $\mathbf{P} = (E_0/c^2) \mathbf{U}$ \cite{Ohanian2009,Ohanian2012, OhanianEinsteinMistakes}. In short, the Laue-Klein Theorem establishes a structural similarity between the energy-momentum of a far more generic class of systems, and that of a point-particle, with rest-energy $E_0$ ($\times c^{-2}$) playing the same role in both---\emph{viz}., that of mass.    

The subtle, but crucial bearing of the \textit{Laue-Giulini Theorem} on the energy-mass relation follows from three simple corollaries.
First, while the Lorentz-boosted system differs in its energy-momentum, $\mathbf{P}\not\equiv \mathbf{P'}= \Lambda \cdot\mathbf{P}$, thanks to $\Lambda^T \mathbf{\eta} \Lambda \equiv \mathbf{\eta}$ the energy-momentum's norm square remains unchanged: $(\Lambda \cdot\mathbf{P})\cdot (\Lambda \cdot\mathbf{P})\equiv \mathbf{P} \cdot\mathbf{P}$. Hence, the boosted system has the same rest-energy as the original system. 
Secondly, by dint of the relativity principle, let's change the perspective, switching from boosted copies of a system, relative to the same observer, to \textit{one} system relative to \textit{different}, Lorentz-boosted observers. Then, their simultaneity hypersurfaces will slice differently through the system; the observers will survey it from different angles. As per the preceding remark, energy-momenta of the system differ merely by a Lorentz transformation, $\mathbf{P'}=\Lambda \cdot\mathbf{P}$, and hence have the same rest-energy.
Thirdly, combining both points with the theorem's key assertion, we arrive at a profound result: the energy-momentum of a system of Laue-Giulini type appears to all Lorentz-boosted observers as the same: it takes the form of point-particle's energy-momentum of the mass $E_0/c^2$, irrespective of the observer. In other words, the energy-momentum of an otherwise arbitrary system doesn't depend on the system's internal structure (information about which one would expect different observers to pick up); no matter the details of the system's composition, its energy-momentum \textit{invariably} is that of a point-particle's with the mass $E_0/c^2$. This absence of internal structure is the distinctive trait of point-particles. The Laue-Giulini version thus further strengthens (under slightly different conditions) the structural similarity of a system's and a point-particle's energy-momentum, with rest-energy, $E_0/c^2$, playing the role of mass in the case of the former.

\subsubsection{The Centre-of-Mass Theorem}

The final context pertinent to (MASS FUNCTIONALISM) is the Centre-of-Mass Theorem (\textbf{§2.3.2}). In a precise sense, it permits us to model blobs of field excitations, continuous fluids or solid matter as point-particles \textit{as far as their centre-of-energy is concerned}.

According to the main result (in the simplified, not fully covariant form, sufficient for our purposes) we obtained Eq. (\ref{CENTRE-OF-MASS}), equivalent to:
\begin{equation}
\frac{E}{c^2} \frac{dX^m}{dt} = P^m = const.
\label{CoM Motion}
\end{equation}

Rewriting this in the centre-of-energy's proper time, $d\tau = (\mathcal{M}c^2/E) dt$ (with $\mathcal{M}^2=\mathcal{M}^2[P] = \mathbf{P}^2/c^2$, which is conserved), we get:

\begin{equation}
\mathcal{M}\frac{dX^\mu}{d\tau} = P^\mu.
\label{centre}
\end{equation}

For time-like $\mathbf{P}$, we already saw that $\mathcal{M}[P]= E_0/c^2$ for the system's rest-energy $E_0$.

Thus, the centre-of-energy, $(X^m)$  moves like a free point-particle with constant momentum, where the latter is given by the centre-of-energy's velocity $dX^b/d\tau$ times $E_0/c^2$. In Eq. (\ref{centre}), rest-energy instantiates the role of mass as the \textit{ratio between momentum and velocity}, at the heart of Newton's First Law in one of its standard forms.

The argument can be further strengthened by relaxing the assumption of the system's isolation (via splitting off a suitable ``external'' $T^{\mu \nu}_{ext}$): under the influence of an external force $\mathbf{F}_{\mathbf{ext}}$ acting on the system, the Centre-of-Mass Theorem reproduces the analogue of Newton's Second Law:

\begin{equation}
(E_0 /c^2)\frac{d^2X^\mu}{d\tau^2} = \mathbf{F}_{\mathbf{ext}}.
\label{centre-SEC-LAW}
\end{equation} 

Again, rest-energy in the form of $\mathcal{M}=E_0/c^2$ instantiates a paradigmatic functional role of mass: that of \textit{inertial resistance} in Eq. (\ref{centre-SEC-LAW}).


A corollary of the foregoing directly bears on how the mass of everyday matter arises. As our paper focuses primarily on mass in classical (as opposed to quantum) field theory, we confine ourselves to cursory remarks on this intricate subject (see \cite{QuarkMassStressEnergy,QuarkMassStressEnergyEncyclopedia,QuarkMassStressEnergyPopular} for details).

Most of the mass effectively probed in ordinary matter resides in protons and neutrons; the contribution of electrons is negligible by comparison. (For the present discussion we may also set aside nuclear binding energy. Conceptually, it's part of the same rest-energy accounting below, but its contribution is likewise comparatively small.) Protons and neutrons aren’t fundamental particles, however. Instead, they are composite states of the quantum fields within quantum chromodynamics (QCD), consisting predominantly of up and down quarks together with gluonic fields. The quark fields possess mass terms (in the sense of \textbf{§2.2}---recall the warning in fn. 9 against na\"{\i}vely identifying these terms with effective masses); the gluon field does not. The strongly interacting QCD environment then generates what is often heuristically described as an effective constituent-quark mass; one may loosely attribute it to a quark inside a proton. The details would require more extensive engagement with QCD; they shan’t detain us here. 

Most germane to our present purpose is that those explicit quark mass terms account for only a tiny fraction of the proton and neutron masses---of order 1\%. The overwhelming remainder arises from the energy of the confined quark and gluon fields and from quantum effects (associated with the dynamically generated QCD scale, in particular the QCD trace anomaly). The mass effectively probed in ordinary matter---the effective mass appearing in the Centre-of-Mass Theorem---is determined by this total energy-momentum of the relevant interacting fields; its energy in the centre-of-mass frame is the system's rest energy, which plays the role of mass. Remarkably, even if---counterfactually--- the up, down, and strange quark masses were set to zero, QCD would still generate a proton mass of roughly 90\% of its actual value. The upshot is that most of the mass of ordinary matter emerges as an energetic consequence of QCD dynamics: the energy stored in the interacting quantum fields makes up the lion's share of the system's rest energy, and hence of the effective masses of everyday matter.



By way of summary, then, \textbf{§3.1.1}-\textbf{§3.1.4} compiled four variegated theoretical contexts (and respective domains of validity). They reveal a close connection between rest-energy and mass in its traditional, mechanical sense:    

\begin{itemize}
    \item a strict correlation between mass and rest-energy ($\times c^{-2}$) for limiting cases,
    \item rest-energy ($\times c^{-2}$) as the proportionality factor between  momentum and acceleration (as per the Centre-of-Mass Theorem and the Euler Equations for pressureless fluids),
    \item rest-energy ($\times c^{-2}$) as the proportionality factor between momentum and velocity (ditto),
    \item the structural similarity---sameness of form---between the energy-momentum of a point particle, and that of an extended system, with rest-energy ($\times c^{-2}$) as the corresponding particle's mass (as per the Laue Theorems).
\end{itemize}

In light of these ties, (MASS FUNCTIONALISM) encourages us to hail rest-energy as the successor term of mass in field theory---as its conceptual generalisation: in pertinent regards rest-energy instantiates the traditional roles that mass has played in mechanics, while also transcending, and eventually superseding mass, as a notion \emph{sui generis}. 

\subsection{Discussion}

Let's further unpack how the two components of the mooted field theoretic interpretation---(ELIMINATIVISM) and (MASS FUNCTIONALISM)---work in tandem. We'll zoom in on four features: its relation to matter or interaction theories (\textbf{§3.2.1}), the compatibility of (ELIMINATIVISM) and the qualification as an interpretation proper of mass and its relation to energy (\textbf{§3.2.2}), its generality (\textbf{§3.2.3}), and the extent to which this relation asserts a form of energy-mass equivalence (\textbf{§3.2.4}).

\subsubsection{Field theory as a \textit{framework} for relativistic \textit{theories}}

As per (ELIMINATIVISM), mass loses its status as a central notion. The field-theoretic energy-mass equivalence relation doesn't fundamentally require inertial mass as an intrinsic property that all material bodies necessarily possess, as prerequisites for momenta and forces (subject to Newton's Laws). Also metaphysically, adherents of the Newtonian framework tend to attach significance to mass (see, \emph{e.g.}, \cite[ch. 1]{Fernflores2017} for a historically more nuanced review): as the quantity of matter itself, it's declared an essential attribute of physical stuff. In field theory, one drops both axioms.

Whether inertial mass in a more standard, mechanical sense enters the picture is a \textit{contingent} matter, beyond SR's jurisdiction. Its answer hinges on further---typically more fundamental---theories that specify the nature of the fields in question. Like Newtonian physics (for, say, Newton's universal law of gravity), field theory is a \textit{framework} for relativistic theories, rather than a relativistic theory itself (such as Maxwellian electrodynamics). It operates at a higher level of abstraction (cf. \cite[ch. 2.6 \& ch. 4, Introduction]{Bunge1967}; \cite{Curiel2021Framework} for similar perspectives). Such theories can instantiate (or ``respect'') a framework. 

Field theory enunciates assumptions of great generality (\textbf{§3.2.3}). In particular, it delimits overarching \textit{constraints} on hypotheses about physical processes (\emph{e.g.}, Yukawa's proposal for proton-neutron interactions) or models of concrete phenomena (\emph{e.g.}, $\beta$-decay).\footnote{Einstein (1919) himself extolled this function, the concomitant generality and its top-down, more axiomatic flair with his characterisation of SR as a ``principle theory'' \cite{Einstein1919Times} (see also, \emph{e.g.}, \cite{Flores1999-FLOETO,BrownPhysicalRelativity,VANCAMP201123,Lange2014-LANDER,Giovanelli2014-GIOBOM} for the philosophical reception of Einstein's distinction). Matter-theoretic models, conforming to the constraints of SR, by contrast, qualify as ``\textit{constructive} theories''. With them, one strives to build a model of the phenomena in a more bottom-up manner.}

As a corollary, this addresses one of the ``main philosophical questions'' concerning the energy-mass relation \cite[sect. 2]{Fernflores2019}: to what extent can mass and energy be \textit{converted into each other}? Given the  inherent conceptual resources of field theory, there is \textit{only} energy-momentum (see, \emph{e.g.}, \cite[ch. 7]{TaylorWheeler}). Strictly speaking, questions about mass and its conversion are alien \textit{to that framework}.\footnote{This is our irenic resolution of the strife between the Conversion and Non-Conversion Interpretations, as \cite[sect. 2.3]{Fernflores2019} calls them. In the main, we side with Bondi and Spurgin \cite{Bondi_1987} in that SR (i.e. relativistic field theory) \textit{as a framework} only traffics in energy-momentum and the transformation of its various forms. What Bondi and Spurgin leave out is what, \emph{e.g.}, \cite[ch. 27]{RIndler1982} correctly stresses: not only does SR not prohibit mass-energy conversions; also some processes are most naturally construed as involving such conversions---processes splendidly described via \textit{special-relativistic theories}. The two parties talk past each other, led astray by an ambiguity of what each means by ``SR'': Bondi and Spurgin's claims concern SR as a framework, whereas Rindler's concern special-relativistic matter theories (within that framework).} 

\textit{How} energy-momentum is realised, especially at a more fundamental level---and, in particular, whether mass in a recognisable form figures at that level and, if so, whether it can be converted into other forms of energy-momentum---is delegated to elementary particle physics. Usually, the coveted matter or interaction theories are, of course, deliberately constructed to respect the constraints from field theory (see, \emph{e.g.}, the classic \cite{bjorken1965}). At present (cf. \cite[sect. 2.3]{Fernflores2019}, the empirically corroborated answer that particle physics furnishes to the question of mass-energy conversions is affirmative. The mass defect of atomic nuclei is a well-rehearsed example: if a nucleus is split, its parts can have a smaller or greater mass than the whole; the difference in mass corresponds to the difference in binding energy. More dramatically, in so-called annihilation/creation processes, particles whose standard physical description posits mass can decay into, or emerge from, fields whose description no longer seems to involve mass (see, \emph{e.g.}, \cite[ch. 8]{TaylorWheeler} for illustrations).

\subsubsection{A \textit{genuine} interpretation of the mass-energy relation?}

(ELIMINATIVISM) is apt to elicit disquiet about the status of the field-theoretic perspective as a genuine interpretation of the mass-energy relation: if the field-theoretic interpretation \textit{isn't really} about mass---but about energy-momentum and rest-energy---doesn't it border on a sleight of hand to tout it as illuminating the mass-energy relation? The latter, after all, is about \textit{mass}.
Two kinds of considerations of that ilk may nourish doubts about the interpretation's legitimacy.

The first attacks the reasoning behind (MASS FUNCTIONALISM) as analogical, and \textit{ipso facto} specious. The Laue Theorems, the analysis of the relativistic Euler Equation, and the Centre-of-Mass Theorem all pivot on structural similarities with the energy-momentum of a point-particle.  One might therefore complain that (MASS FUNCTIONALISM) capitalises on a formal analogy (like that, say, between the Black-Scholes equation for financial markets and the heat equation for the diffusion of heat). It educes from the same structural relations amongst momenta, velocities, accelerations and external forces some (attenuated) kind of identity about mass and rest-energy. Such an argument, one might protest, is neither deductively valid, nor  inductively persuasive (as attested by the abundance of formal analogies in science). What assures us that the analogy is no mere coincidence? 

Secondly, reservations are further aggravated by the restrictive conditions that each of the three theorems requires (and \emph{a fortiori}, the highly idealised limiting cases). Laue's Theorem (especially in its original form and the Laue-Giulini form) and the pressure-free relativistic Euler Equation rule out a large array of realistic scenarios. 

Moreover, upon a little reflection, the structural relations themselves between momentum and velocity, and force and acceleration, are less general than one may initially have hoped. Recall, for instance, that mass can \textit{cease to be the proportionality factor} between velocities and momentum already within classical mechanics (see, \emph{e.g.}, \cite[sect. 7]{Landau1976}) for coordinates other than Cartesian ones (\emph{e.g.}, polar ones), or for velocity-dependent interaction terms in a Lagrangian (as for a charged particle in an electromagnetic field). 

Should we conclude, as the objection insinuates, that rest-energy only imitates mass? Does field theory, with its energy-momentum, proffer a \textit{simulacrum} of mass, which merely exhibits a flimsy similarity, devoid of deeper physical significance? Such a conclusion, we would counter, underestimates the thrust of (MASS FUNCTIONALISM):

\begin{itemize}

\item The limiting cases moor rest-energy in bona fide mass of arguably the most paradigmatic physical object in a physicist's training.\footnote{In a sense the prototype of mass, in its point-particle mechanical form, needn't even be postulated independently: certain conditions on the energy-momentum tensor alone \textit{entail} the equations of motion for massive particles \cite{Geroch2018,Weatherall2018geometrymotiongeneralrelativity}.} Admittedly, this connection is too weak to underwrite claims of uniqueness. Yet, as a conceptual bridge, it shouldn't be underrated either.  
\item Despite lack of \textit{unadulterated} generality, the relativistic Euler Equation, the Centre-of-Mass Theorem, and Laue's Theorem employ the two putatively \textit{most salient}\footnote{For instance, Einstein's \cite[ch. 1]{Einstein1922Meaning} or Born's \cite[ch. II.9\&10]{Born1922}) exposition of Newtonian Mechanics uses the mass-velocity-momentum and the mass-acceleration-force relation (likewise, \cite[ch. 37]{Pauli}} roles of mass. 
\item The theorems invoked in \textbf{§3.1} hold under different circumstances. The variety and number of functional links between rest-energy and mass lend the network of similarity relations robustness.
\end{itemize}
 Notwithstanding its imperfections\footnote{Such dissimilarities and mismatches routinely occur in inter-theory relations---grist for the mills of those in favour of more liberal, ``imperfection-tolerant'' (rather than rigorously formalistic) approaches. An instructive example is the ``reduction'' of General Relativity to Newtonian Gravity (\emph{e.g.}, \cite[section 5]{KnoxWallace}, who portray the case as an illustration of what they call ``constitutive functionalism''---a stance similar in spirit to the strategy of (MASS FUNCTIONALISM).}, these three points make the functional family resemblance between mass and rest-energy, to our minds, sufficiently strong and non-arbitrary. It justifies pronouncing rest-energy the successor concept of mass (with the semantic-conceptual counterpart of ``Kuhn losses''): rest-energy qualifies as its field-theoretic \textit{generalisation}.\footnote{Kuhn has forcefully emphasised family resemblance (rather than sharp, jointly sufficient and/or necessary conditions) as the primary avenue for learning new classifications, concepts, and problem-solving techniques (see \cite[ch. III]{Hoyningen-Huene1993-HOYRSR}. Other examples of such a (semantically non-monotonic) generalisation of a concept where family resemblance anchors the successor term (despite semantic-conceptual discontinuities) comprise the concept of a particle in modern particle physics \cite{falkenburg2007particle} or black hole entropy \cite[section 5]{sep-spacetime-singularities}. An especially compelling example from outside of physics is arguably that of a gene (see, \emph{e.g.}, \cite{portin1993gene}.}

\subsubsection{Generality}

The double sense of generalisation deserves to be highlighted. First, the ``classical'' proofs of the energy-mass relation establish it for various scenarios (\emph{e.g.}, Einstein's radiation-filled box, or the photon-emitting atom) in  \textit{conditional} form: one shows that consistency with a mechanical description demands that a system's effective mass and its rest-energy be related. That is, they demonstrate that, \textit{if} we assume the system to be subject to the laws of relativistic mechanics (including its principles of energy and momentum conservation), the bulk mass entering the mechanical laws must be $E_0/c^2$, with the system's rest-energy $E_0$.


The field-theoretic theorems of \textbf{§3.1.2}-\textbf{§3.1.4} ``de-conditionalise'' the previously conditional claim. They prove that modelling a system ``mechanically'' is kosher: the bulk system's energy-momentum \textit{indeed} behaves like---may be treated as---a mechanical system. We needn't presuppose the laws of relativistic mechanics \emph{ab initio}.\footnote{Such modelling tends to operate with rigid bodies, which fact compromises the status of the arguments: as it dawned on Einstein as early as 1907, the concept of rigidity sits uncomfortably within any special-relativistic theory (see, \emph{e.g.}, \cite{giovanelli2023length}).} Rather, one \textit{derives} the structural analogy with a relativistic-mechanical system. 

This result segues into the second sense of generalisation that the field-theoretic interpretation accomplishes: the derivations in those theorems utilise only the more abstract elements of relativistic field theory, rather than more specific posits of relativistic matter theories. This seems fitting for  ``the most important upshot of the special theory of relativity'' \cite[p. 230]{Einstein1919Times}.


\subsubsection{The energy-mass relation and mass-energy \textit{equivalence}?}

Finally, let's broach whether the energy-mass relation asserts an ``equivalence'' between mass and energy---as the common appellation of the energy-mass relation suggests. 
Call two properties (logically) equivalent if they have the same truth-conditions for their ascription. With equivalence thus construed, the field-theoretic interpretation repudiates the popular label for the energy-mass relation: insofar as talk of equivalence implies a symmetric relation between energy \textit{simpliciter} and mass, relativistic field theory \textit{gainsays} an energy-mass equivalence.

An equivalence thesis about \textit{rest}-energy---as the relevant form of energy the energy-mass relation is concerned with---fares no better. In line with (ELIMINATIVISM), within the framework of field theory, mass isn't a fundamental or essential property of matter, whereas energy-momentum is. Mass is a property that certain relativistic theories may---or may not---posit. Energy-momentum, by contrast, is defined for every field (subject to isolation, typically codified through suitable fall-off conditions).


Furthermore, according to the core tenet of the field-theoretic interpretation, rest-energy counts as the successor term of mass in field theory: in line with (MASS FUNCTIONALISM), it instantiates salient aspects of the functional profile of mass under certain circumstances. At the same time, (MASS FUNCTIONALISM) concedes that this functional family resemblance is far from perfect or universal.  

These asymmetries vitiate claims about rest-energy-mass \textit{equivalence}; despite usage in ``many known texts by W. Pauli, P.G. Bergmann, C. M{\o}ller, E.F. Taylor and J.A. Wheeler, [etc.]'' \cite[p. 88]{Jammer2000}, the popular slogan, on the field-theoretic interpretation, is a misnomer.\footnote{Bunge pithily judges: ``(t)he numerical equivalence of $m$ and [rest-energy] holds only for systems that are assigned a mass to begin with; and strictly speaking it is as little an equivalence as the linear relationship between force and displacement in Hooke's law. [...] In general: numerical equalities do not entail the identity of the predicates involved'' \cite[p. 201]{Bunge1967}.}


\section{Unifying power and coherence of the field-theoretic interpretation}

In our preceding discussion several merits of the proposed interpretation already shone through. Here, we'll scrutinise them more systematically. These virtues form a coherent cluster, instructively parsed into aspects of unification. Following Bartelborth \cite{Bartelborth2002}\cite[ch. VI]{Bartelborth2007}, let's distinguish amongst three different dimensions of unifying power (of a theory, model, or classificatory system):   
\begin{itemize}
\item \textit{Empirical scope} How numerous, and how variegated are its empirical applications, or the diversity of phenomena that can be subsumed? 
\item \textit{Systematicity and internal structure:} How rich and strong are the explanatory links and general coherence of postulated principles? Things are supposed to hang together as tightly as possible in a way that bland disjunctions don't. Such connections generate an organic structure and cohesiveness. They enable transfer of information and powerful scientific inferences.
\item \textit{Specificity and information content:} How specific and demanding are the theoretical and conceptual structures in question? What and how much do they prohibit? How stringent are the constraints imposed on possible applications?
\end{itemize}
The field-theoretic interpretation scores high on all three dimensions. 

\subsection{Scope}

As an abstract framework (\textbf{§3.2}), field theory is marked by its generality. It doesn't prescribe exhaustive details for specific types of matter and their interaction. Instead, it delineates broad constraints that more specific theories must respect, and general principles that are supposed to guide research leading to such theories. Field theory purports to subsume all types of (non-quantum) matter. It covers---and extrapolates---established classical physics and its constitutive principles, with electromagnetism, fluid dynamics, and general relativity as its exemplars (\emph{e.g.}, \cite{Landau}. It even covers the wave equations of relativistic quantum mechanics, such as the Klein-Gordon or the Dirac Equation (\emph{e.g.}, \cite[ch. 7]{Goenner2004}).\footnote{As far as their structure qua wave equations is concerned, they count as classical (see, \emph{e.g.}, \cite[section 2]{KantParticle}). It's their physical motivation and interpretation that make them quantum physical.} This generality is bequeathed to the field-theoretic interpretation of the energy-mass relation. After all, the field-theoretic interpretation boils down to a judicious at-face-value reading of field theory; (ELIMINATIVISM) and (MASS FUNCTIONALISM) merely flesh out the status of mass and rest-energy. 

The field-theoretic interpretation can also boast about broad scope in its own right. In contrast to virtually all standard demonstrations of the energy-mass relation, both the interpretation itself and the arguments on which it rests are emancipated from any specific physical setups or interactions (see \cite[ch. 3]{Fernflores2017} \cite[sect. 3]{Fernflores2019}).\footnote{For instance, Einstein's \cite{Einstein1905} (cf. \cite[5.2]{Norton2014_EinsteinSpecialRelativity}  simplification) treatment deals with the \textit{emission/absorption of light quanta}; his 1906 follow-up (see also \cite[ch. Vii.8]{Born1922}) deals with a cavity---presumably within a rigid body!---filled with \textit{electromagnetic radiation}, and his 1935 derivation deals with \textit{inelastic collisions of particles}.}     

\subsection{Systematicity}
The field-theoretic interpretation of the energy-mass relation weaves a rich tapestry of connections and structural relations that flow directly from field theory. They shed light on the treatment of photons (\textbf{§4.2.1}), the synthesis of mass, energy, and momentum (\textbf{§4.2.2}), as well as on a unification of conservation laws for ``our two concepts of substance'' \cite[pp. 54, 208]{Einstein1938evolution}(\textbf{§4.2.3}).

\subsubsection{A coherent integration of photons}

A first benefit in terms of systematicity is that field-theoretic interpretation accommodates \textit{both} photons (or massless particles, more generally, such as gluons\footnote{The existence of reasonable models of relativistic \textit{continuous} matter with no rest mass further illustrates how mass isn’t a central concept in field theories. One can also write down models of more nearly mechanical systems, ``null dust''  or (allowing pressure) a ``null fluid'' with energy density, momentum density, and their currents, with no rest mass.  These continuum models (which can obtained as ultra-relativistic limits of relativistic dust or fluids, \cite{BicakKucharNullDust}) involve matter travelling at the speed of light. Although the world doesn’t seem to contain anything that is fundamentally a null fluid, such a description is sometimes a serviceable description for certain kinds of radiation.}) and massive particles.

With respect to mass, photons appear to defy a coherent interpretation. They spawn three \textit{prima facie} paradoxes:  

\begin{enumerate}[label=(P\arabic*)]
    \item On the one hand, the mass one would be inclined to assign them (from Eq. (\ref{MOMENTUM-ENERGY-3momentum})) is zero. On the other hand, photons incontrovertibly carry energy. A \textit{na\"{i}ve} application of $E=mc^2$ would imply a non-zero mass. How to defuse this apparent tension?
    
    \item A system of photons, it seems (see, \emph{e.g.}, \cite[p. 232]{TaylorWheeler}), \textit{can} have non-vanishing mass, even though the individual photons don't. How does one account for this bewildering ``materialisation'' of mass? Moreover, how should we comprehend the fact that the mass of a swarm of photons depends on the angles of their trajectories?
    
    \item Quite generally, the notion of mass for photons remains rather opaque: at least outside of quantum electrodynamics, forces aren't exerted on them in any straightforward sense (though one could perhaps take the view that gravity according to General Relativity does so). So what does one mean by ``mass of a photon''?
\end{enumerate}

For a coherent field-theoretic perspective on photons, we only need to recall that, like any energy-momentum, a photon's energy-momentum $\mathbf{P}=(P^\mu)$ (or the energy-momentum of a composite system made up of photons) can be uniquely decomposed into temporal and spatial parts relative to an observer (eq. \ref{DECOMPOSE})). For \textit{individual} photons (see, \emph{e.g.}, \cite[ch. 33]{RIndler1982} \cite[ch. 9.2.4]{Gourgoulhon2013}), $\mathbf{\xi} = (c, \vec{n})$ must be light-like, $\boldsymbol{\xi} ^2 =0$; its energy-momentum has the form: $\mathbf{P} = (E/c^2)\boldsymbol{\xi}$. Consequently, by Eq. (\ref{MOMENTUM-ENERGY-3momentum}), $\mathbf{P}^2 \equiv 0$. 

An elegant field-theoretic stance towards the mass of photons ensues:
\begin{itemize}
    \item Individual photons possess non-trivial energy-momentum: $\textbf{P}=(E/c^2)\mathbf{\xi}$. 
    \item It's light-like: $\mathcal{M}^2[P]/c^2=\textbf{P}^2=0$ (with the \textit{definition} (\ref{NORM})). 
    \item By the same token, $\textbf{P}$ has no rest frame. Hence, photons have no rest-energy. (NB: We deny the ascription of rest-energy to photons. This \textit{isn't} the same as to ascribe them zero rest-energy, see, \emph{e.g.}, \cite{Balashov1999}. Photons cannot be at rest.)
    \item Not being natural objects within relativistic mechanics, photons possess no mass (in the sense of mechanics). Not possessing rest-energy, individual photons don't possess the field-theoretic successor term of mass---as per (MASS FUNCTIONALISM)---either: $\mathcal{M}$ doesn't instantiate any of the salient functional roles of mass.
\end{itemize}

These interpretative principles resolve the three paradoxes above. The tension in (P1) arises from two misconceptions. First, a photon doesn't possess a zero-valued mass (\emph{pace}, \emph{e.g.}, \cite[p. 230]{TaylorWheeler}  p.230 \cite[p. 272]{Gourgoulhon2013}); it has \textit{no} mass. (Here we have in mind the literal mechanics sense of mass, not the field theoretic dispersion relation sense of mass relating frequency and wavelength, which we have called figurative.)   
Secondly, the na\"ive application of the energy-mass relation is doubly flawed. The energy that figures in $E=mc^2$ is, as underlined in \textbf{§3.1}, \textit{rest}-energy---not energy simpliciter. Even so, a photon doesn't possess rest-energy. The energy-mass relation becomes \textit{inapplicable}. 
A similar rectification explains away (P3): the mystery evaporates for photons possess neither mass nor mass-like rest-energy.

At first glance, the field-theoretic interpretation exacerbates the mystery  enshrouding (P2): a novel property, mass, appears to emerge in \textit{composite} photonic systems from massless constituents (\emph{c.f.} \cite{Benitez2022}); from the austere perspective of field theory, however, mass doesn't emerge \emph{ex nihilo} as a qualitative novelty of composite systems, because it doesn't emerge at all! To dispel the mystery, first recall a trivial lesson from vector addition in linear algebra (see, \emph{e.g.}, \cite[part V]{RIndler1982}): the norm of a composite system's energy-momentum, $\mathbf{P}_{A\&B} \equiv (\mathbf{P}_A+\mathbf{P}_B)$, differs from that of the sum of its constituent parts, $\mathbf{P}_A$ and $\mathbf{P}_B$): $\mathbf{P}_{A\&B}^2 \neq \mathbf{P}_A^2 +\mathbf{P}_B^2$. 

Next, recall that in line with (ELIMINATIVISM), energy-momentum is the central concept; it's not essential that physical systems possess mass. What the field-theoretic interpretation asserts, with its commitment to (MASS FUNCTIONALISM), is that, whenever rest-energy exists, it bears a functional family resemblance to mass, in virtue of which rest-energy qualifies as the successor concept of mass. Neither individual photons nor composite photon systems possess mass \textit{sensu stricto}. Mass thus doesn't materialise or emerge from zero-mass or massless constituents. Instead, depending on the (accidental) composition of the system, the energy-momentum of a composite system need no longer be light-like. If it's not, the system possesses a non-zero rest-energy. In some contexts, it in turn, according to (MASS FUNCTIONALISM), plays a functional role similar to---but \textit{not identical} to---mass. Like (P1) and (P3), the paradox (P2) is debunked as fallaciously presupposing the literal ascription of mass, which (ELIMINATIVISM) blocks.



\subsubsection{Synthesis of mass, energy, and momentum}

An especially intriguing achievement of the field-theoretic interpretation lies in its ability to unify energy, momentum, and (to some extent) mass. 
In pre-relativistic physics, energy and momentum are distinct quantities. Relativistic field theory, by contrast, fuses them: energy-momentum or, in its local form, energy-stress represents \textit{both} through one object, a four-vector or a rank-2 tensor field, respectively. They are linked via new law-like connections. Through Lorentz boosts the components of the energy-momentum vector \textit{mix} (with important applications, for instance, in the astrophysics of jets ejected from the vicinity of black holes, see, \emph{e.g.}, \cite[ch. 21.7]{Gourgoulhon2013} for details). Consequently, energy and momentum are profoundly intertwined---in a way completely analogous to the way the electric and magnetic field are unified in electrodynamics (see, \emph{e.g.}, \cite{Maudlin1996-MAUOTU-2}\cite[section 2]{Morrison2013}): they forfeit their independence and fundamental distinctness. As Eq. (\ref{DECOMPOSE}) made explicit, energy and momentum become frame-relative---time-like and space-like---projections. They earn Minkowski's  poetic paraphrase \cite[p. 104, our translation] {MinkowskiSpaceTimeShadows}, just as space and time do in their spatiotemporal \textit{Aufhebung}: in SR, energy and momentum ``fully sink into mere shadows'' of the superordinate entity ``energy-momentum'' \cite[Ch. 7]{TaylorWheeler} \cite[Ch. 9.2.2]{Gourgoulhon2013}, ``and only some sort of union of the two shall preserve autonomy''. 

Physical descriptions must therefore be formulated in terms of the entity that casts those shadows: namely energy-momentum. This is the insight that, in his Princeton lectures, Einstein (as cited in \textbf{§1}) accentuated. A direct (and historically relevant, see \cite{NortonThoughtLesser,Norton}) application is relativistic hydrodynamics (see, \emph{e.g.}, (\cite[Ch. 4.7.1]{Goenner2004} \cite[Ch. 21]{Gourgoulhon2013}). As a ramification, systematicity is further amplified: pressure, shear and energy density are likewise unified in that they are constructed from the energy-momentum tensor. The subsumption of hydrodynamical quantities under the energy-momentum tensor with its novel structural links also results in several new effects, such as the contribution of pressure to the dynamics of a fluid (more on this in \textbf{§4.3}). 


Finally, recall that energy-momentum is, as we saw, often related to mass. Mass or mass densities are postulated as parameters or variables in continuum mechanics (a posit at the level of a specific relativistic matter theory, see \emph{e.g.}, ibid.). Through their presence in energy-momentum complexes, they contribute to the system's energy-momentum. In this sense, the field-theoretic unification also includes mass.\footnote{Arguably, the kind of unification of mass and energy-momentum differs from that of energy and momentum: the latter is an instance of what Morrison calls ``reductive unification'' \cite{Morrison2013}, whereas the former is a case of ``synthetic unification''.}  


\subsubsection{Unification of conservation laws}

A last element of systematicity which commends the field-theoretic interpretation pertains to conservation laws---its parsimony in that respect. Einstein himself averred that ``(t)he most important result of a general character to which the special theory of relativity has led is concerned with the conception of mass. Before the advent of relativity, physics recognised two conservation laws of fundamental importance, namely, the law of the conservation of energy and the law of the conservation of mass; these two fundamental laws appeared to be quite independent of each other. By means of the theory of relativity they have been united into one law'' \cite[p. 60]{Einstein1916relativity} (see also, \emph{e.g.}, \cite[pp. 208, 259]{Einstein1938evolution}).


Einstein's gloss fits the field-theoretic interpretation like a glove. In line with its (ELIMINATIVISM), the relevant conservation law must be couched in terms of energy-momentum, not in terms of mass. We discussed its conservation law in \textbf{§2.3}. SR \textit{on its own} doesn't ensure mass conservation. Particle physics allows for the (empirically confirmed) creation and annihilation of certain types of (massive) particles, for instance, pairs of electrons and positrons that annihilate each other, producing photons. Mass conservation is thus violated.


\subsection{Specificity and informativeness}

Its specificity safeguards the field-theoretic interpretation's sizeable unifying power. One argument to this effect again trickles down from the field-theoretic framework. The gist of relativistic field theory is to urge the incorporation of Lorentz covariance (see \cite[ch.4 section 4]{BrownPhysicalRelativity}): the demand that \textit{all} of physics remain invariant under Lorentz transformations.\footnote{More precisely, physical quantities must be representations of the Lorentz group $\text{SO}(1,3)$, (or, if we include also invariance under translations $\mathbb{R}^{1,3}$, the Poincaré group $\mathcal{P} = \text{SO}(1,3) \ltimes \mathbb{R}^{1,3}$) see, \emph{e.g.}, \cite[ch. 6]{Goenner2004} or \cite[chs. 2, 3]{Maggiore2005qft}. For the ``spinor'' fields describing half-integral spin particles such as electrons, neutrinos, and quarks, a slight generalisation is neeeded.}
Such a universal postulate is indubitably a bold claim.\footnote{It's also worth mentioning that Minkowski spacetime, as the spacetime with the Lorentz group, $\text{SO(1,3)}$, as the global isometry group, is uniquely distinguished by its \textit{mathematical simplicity}: discarding de Sitter and anti-de Sitter spacetime (associated with a non-vanishing cosmological constant), it's the simplest of the three maximally symmetric spacetimes. Insofar as symmetries typically enhance a theory's content---by ruling out possibilities---the Minkowskian spacetime is a highly specific/contenful posit.} Correlatively, it's eminently testable (see, \emph{e.g.}, \cite{Mattingly2005} for a review). It thus has substantive---in fact, \textit{well-corroborated}---content. Insofar as the field-theoretic interpretation of the energy-mass relation solely draws on the conceptual resources of field theory, it partakes of the latter's specificity.


Secondly, and relatedly, the field-theoretic interpretation may legitimately be credited with (successful) predictions, at least qualitative ones---another hallmark of specificity of content. Being effectively limited to particle mechanics, most other interpretations are condemned to silence or vagueness on physics beyond that turf. One such prediction is, as Einstein discerned, that we should expect a radiation-filled or stressed body to gravitate more. But also more specifically, if energy-momentum is the fundamental quantity that replaces mass, it follows that field equations for a relativistic theory of gravity must be constructed from the energy-momentum tensor as a source term---the path leading to General Relativity (\cite{Norton}).

\section{Summary and conclusion}
The paper propounded a thoroughgoing field-theoretic interpretation of the energy-mass relation. Relativistic field theory, we maintained, provides the most complete and refined articulation of SR as a theoretical framework---at least as far as \textit{classical/non-quantum} physics is concerned.\footnote{As flagged from the outset, our paper generally is restricted to classical field theory. An important follow-up question, to be tackled in future work, is therefore: to what extent do our results carry over to \textit{quantum} field theory? Already our brief comments on nucleon masses (\textbf{§3.1.4}) intimate that one should expect to see radical conceptual novelties (associated with, e.g., the promotion of tensor fields by operators, the impact of renormalisation on energy decompositions or the role of so-called quantum anomalies) that complicate the quantum field theoretical situation.}

In field theory, energy-momentum supersedes (inertial) mass as an essential property of matter. Energy-momentum is a \emph{sui generis} fundamental quantity that unifies energy, momentum and (to some extent also) mass. Mass, within field theory, is demoted to a more contingent property; relativistic matter/interaction theories may, or may not, posit it. 

The energy-mass relation in its iconic form, $E=mc^2$, clarifies the connection between mass and energy-momentum. The energy that it refers to is \textit{rest}-energy. It bridges energy-momentum and mass in two ways. First, in special and limiting cases, rest-energy $\times c^{-2}$ reduces to mass in the ordinary/mechanical sense. Secondly, in some theoretical contexts, rest-energy instantiates some of the salient functional roles that mass plays in pre-relativistic physics: as the ratio between momentum and velocity, as the relation between momentum and external forces, and in a structural correspondence between a system's effective energy-momentum and that of a point-particle's. 

This grounds a functional family resemblance with mass as traditionally understood. It justifies identifying rest-energy as the successor term (or field-theoretic generalisation) of mass. Somewhat ironically perhaps, our proposed interpretation of the energy-mass relation hence ends up less concerned with mass sensu stricto than with its functional analogue. 
The field-theoretic interpretation's chief appeal lies in its capacity for coherence and unification.\footnote{We'll leave a detailed comparison of the various interpretations (\emph{e.g.}, \cite{Fernflores2017,Fernflores2019,Coffey2021ontology,Benitez2022}) to future work.} 

The field-theoretical interpretation naturally emerges from---and is woven into the fabric of---relativistic field theory. The latter has proven to be an eminently fruitful framework for 20\textsuperscript{th} century gravitational and quantum physics (see, \emph{e.g.}, \cite{Fernflores2017}).\footnote{One can trace back the fruitfulness of field theory to its combination of generality (\textbf{§4}) and tightness of constraints which it imposes on specific matter/interaction theories: it prescribes that physical quantities be representations of $\text{SO}(1,3)$, and that the differential operators of considered Lagrangians or field equations preserve invariance under Lorentz transformations. This affords a rich toolbox for systematically constructing matter or gravitational theories (see, \emph{e.g.}, \cite[ch. 7]{Goenner2004}\cite{KantParticle}).}   
Our interpretation of the energy-mass relation thus embeds it into a broader and more coherent conceptual framework, ``embracing all phenomena of nature'' \cite[p. 209]{Einstein1938evolution}, one that has engendered triumphantly successful, more specific interaction theories, and continues to shape modern physics. This, we judge, well befits what Einstein and others deemed the most important result of SR.  

\section*{Acknowledgements}

The authors gratefully acknowledge insightful feedback and discussion from Yemima Ben-Menahem (Hebrew University of Jerusalem), Hasok Chang (Cambridge), Neil Dewar (Cambridge), Marco Giovanelli (UNITO, Turin), Michel Janssen (Minnesota), Dennis Lehmkuhl (Bonn) and James Read (Oxford). 
Many thanks also to the members of the History and Philosophy of Science reading group at Cambridge! 
 




\end{document}